\documentstyle[twoside,fleqn,espcrc2,epsf]{article}

\newcommand{\AmS}{{\protect\the\textfont2
  A\kern-.1667em\lower.5ex\hbox{M}\kern-.125emS}}

\newcommand{\al}{\alpha}
\newcommand{\bt}{\beta}
\newcommand{\gm}{\gamma}
\newcommand{\dl}{\delta}

\newcommand{\lm}{\lambda}

\newcommand{\vr}{\varphi}

\newcommand{\ps}{\psi}
\newcommand{\om}{\omega}

\newcommand{\Tr}{\mbox{Tr}\,}  

\newcommand{\dmu}{\partial_{\mu}}

\newcommand{\psb}{\bar{\psi}}

\newcommand{\eela}[1]{\label{#1}\end{equation}}
\newcommand{\eeala}[1]{\label{#1}\end{eqnarray}}
\newcommand{\be}{\begin{eqnarray}}
\newcommand{\ee}{\end{eqnarray}}
\newcommand{\bea}{\begin{eqnarray}}
\newcommand{\eea}{\end{eqnarray}}
\title{
\vspace*{-30pt}
{\normalsize \hfill {\sf ITFA-99-23}} \\
\vspace*{-6pt}
{\normalsize \hfill {\sf THU-99-26}} \\
Real-time dynamics in the 1+1 D abelian Higgs model with 
fermions\thanks{Work supported by FOM/NWO, presented by J.~Smit. 
}
}

\author{Gert Aarts\address{Institute for Theoretical Physics, 
        Utrecht University\\
        Princetonplein 5, 3584 CC Utrecht, the Netherlands}
        and 
Jan Smit\address{Institute for Theoretical Physics, 
        University of Amsterdam \\
        Valckenierstraat 65, 1018 XE Amsterdam, the Netherlands}$^{\rm ,a}$
}       
\begin{document}

\begin{abstract}
In approximate dynamical equations, inhomogenous
classical (mean) gauge and Higgs fields are coupled
to quantized fermions. The equations are solved numerically on a
spacetime lattice. The fermions appear to equilibrate 
according to the Fermi-Dirac distribution with time-dependent temperature 
and chemical potential.
\end{abstract}

\maketitle

1. The real-time path integral for quantum fields is very difficult
to evaluate numerically
and approximations need to be made before giving the problem to the computer.
Two types of approximations are currently in use:
classical and gaussian, such as Hartree, large $N$.
The classical approximation 
gives valuable nonperturbative results \cite{Moore}
but suffers complications due to (Rayleigh-Jeans type) divergencies.
The gaussian approximation
has the benefit of staying within the quantum domain where we know how to
deal with divergencies, but it is not good enough for large times.
Naturally, one would like to combine the good aspects of both approximations 
\cite{SB99}.
A crucial test is to see whether the system equilibrates quantum-like, and
not classical equipartition-like, despite the fact that one is just solving
a large number of coupled nonlinear equations which conserve energy.
This is one motivation for the present study. Another is
the intrinsic interest in the complicated nonperturbative
dynamics of the abelian Higgs model with fermions.


2. The 1+1 D abelian Higgs model coupled axially to fermions 
is qualitatively similar to the electroweak sector of the 
Standard Model. As for the SU(2) case in 4D, it can be rewritten in
a form with vectorial gauge couplings and Majorana-Yukawa couplings.
For $N\to \infty$ fermion replicas, the equations
of motion reduce to a classical
field approximation for the bosonic variables,
with a quantal fermion backreaction \cite{AaSm98}:
\bea
\dmu F^{\mu\nu} + e^2 i(D^{\nu}\vr^*\,\vr-\vr^* D^{\nu}\vr)
&&\nonumber\\ \mbox{} 
+ (e^2/2) \langle\bar\ps i\gm^{\nu} \ps \rangle &=&0,
\nonumber
\\
(-D_{\mu}D^{\mu} + \mu^2 + 2\lm\vr^*\vr)\vr &=&0,
\nonumber
\eea
where we have specialized to zero Yukawa coupling.
The fermion backreaction
is specified as follows. Introduce a complete set of orthonormal mode functions
$u_{\al}, v_{\al}$ for the fermions, which satisfy the Dirac equation
\be
\gm^{\mu} D_{\mu} u_{\al} =0,
\;\;\;
\gm^{\mu} D_{\mu} v_{\al} =0.
\nonumber
\ee
Next define the fermion operator $\hat\psi$,
\[ 
\hat\psi(x) = \sum_{\al}[\hat b_{\al} u_{\al}(x) 
+ \hat d_{\al}^{\dagger} v_{\al}(x)],
\nonumber
\]
in terms of annihilation and creation operators $\hat b_{\al}$, 
$\hat b_{\al}^{\dagger}$, \ldots. The fermion back reaction is then
specified by the initial conditions 
$\langle \hat b_{\al}^{\dagger} \hat b_{\al'}\rangle = n_{\al}\, \dl_{\al\al'}$,
$\langle \hat d_{\al}^{\dagger} \hat d_{\al'}\rangle = 
\bar n_{\al}\, \dl_{\al\al'}$,
 etc.

The above system of equations has been implemented on a lattice using
Wilson's fermion method for the spatial derivative and the staggered
fermion
interpretation for a `naive' discrete time derivative \cite{AaSm98}.
Usual expectations on fermion number non-conservation tied to sphaleron
transitions are correctly represented on the lattice.
For simplicity we continue with continuum notation.

\begin{figure}[t]
\epsfxsize 7.9cm
\centerline{\epsfbox{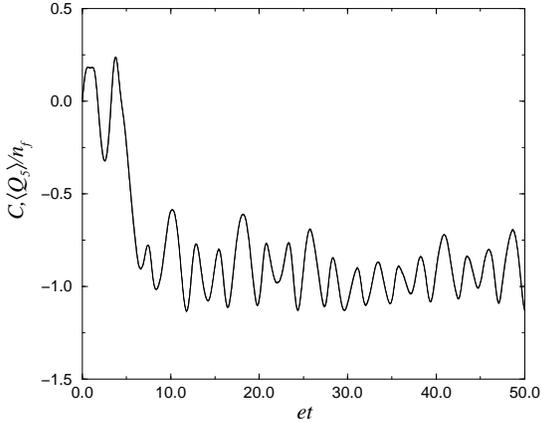}}

\vspace{-1cm}

\caption{Chern-Simons number $C$ and axial charge $\langle Q_5\rangle/n_f$ 
versus time in units $1/e$.
}
\label{fig1}
\end{figure}
\begin{figure}[t]
\epsfxsize 7.9cm
\centerline{\epsfbox{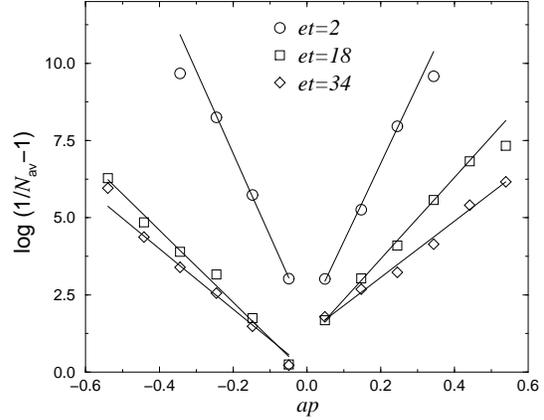}}

\vspace{-1cm}

\caption{Least squares straight line fits to $\ln(N_{{\rm av}p}^{-1} - 1)$ 
versus $ap$. 
}
\label{fig2}
\end{figure}
3. Here we are especially interested in thermalization properties of
the fermions. 
We tested \cite{AaSm99} for local (in time) equilibration of fermions by
comparing their distribution function with the Fermi-Dirac distribution.
The distribution function was identified
from the equal-time fermion two point function, averaged over space (a circle
with circumference $L$), 
\be
S(z,t) = 
\frac{1}{L}\int_0^L dx\,
\langle\ps(x,t)\psb(x+z,t)\rangle_{\mbox{g.f.}}.
\nonumber
\ee
Here g.f.\ indicates a complete gauge fixing.
Alternatively, 
the two point function can be rendered gauge invariant by supplying a parallel
transporter $U(x,y) = \exp[-i\int_x^y dz\, A_1(z)/2]$. We have used the latter
method, but it is actually closely related to complete Coulomb
gauge fixing \cite{AaSm99}. If the fermions were free, the Fourier transform
\be
S(p,t) = \int_0^L dz\, e^{-ipz}\, S(z,t)
\nonumber
\ee
would be given in terms of distribution functions $N_p$, $\bar N_p$ als follows:
\bea
\Tr S(p,t) &=&
[1- N_p(t) - \bar N_{-p}(t)]
\,\frac{m_p(t)}{\om_p(t)},
\nonumber\\
\Tr i\gm^1 S(p,t) &=&
[1- N_p(t) - \bar N_{-p}(t)]\,
\frac{ p}{\om_p(t)},
\nonumber\\
\Tr i\gm^0 S(p,t)&=&
1- N_p(t) + \bar N_{-p}(t),
\nonumber\\
\om_p(t) &=& \sqrt{m_p^2(t) + p^2},
\nonumber
\eea
with $\Tr \gm_5 S(p,t)=0$ because of parity invariance. 
For free fermions $N_p$, $\bar N_p$ and $m$  are time-independent.
Assuming that the interacting model
can be described approximately by quasiparticles, we now use the above
equations to {\em define} $N_p(t)$, $\bar N_p(t)$ and $m_p(t)$. 
In a non-equilibrium
situation they will depend on time.

The simulations had the following parameters: 
$n_f = 2$ flavors (related to fermion doubling in time), 
spatial size $m_{\vr}L \approx 6.4$, 
$\lm/e^2 = 0.25$ ($m_A L \approx 9$), coupling $e^2/m_{\vr}^2 \approx 0.25$,
with spatial lattice spacing $am_{\vr}\approx 0.10$, temporal spacing
$a_0/a = 0.005$
and $L/a = 64$ spatial lattice sites.

Fig.\ \ref{fig1} shows the Chern-Simons number $C = -\int dx\, A_1/2\pi$ and
the axial charge $\langle Q_5\rangle/n_f$ for a simulation 
starting with a fermionic vacuum
(i.e.\ $n_{\al},\bar n_{\al} = 0$) 
and some kinetic energy stored in a few low
momentum modes of the Higgs field. 
The anomalous fermion number non-conservation equation
$\Delta \langle Q_5 \rangle = n_f \Delta C$,
is well obeyed since the two curves are indistinguishable (initially 
$\langle Q_5\rangle = C = 0$).
The oscillations correspond roughly to the basic period $2\pi/m_A$.
To smoothen these we average $S(p,t)$ over a time interval $t_{\rm av}$
before extracting the distribution functions. 
We used $et_{\rm av} = 4$ and studied
the behavior of $N_{{\rm av}p}(t) \equiv [N_p(t) + \bar N_{-p}(t)]/2$.
As expected, $m_p(t)$ is effectively zero. 
Surprisingly, $N_{{\rm av}p}(t)$
resembles quite fast a Fermi-Dirac distribution 
for inverse temperature $\bt$ and $Q_5$-chemical potential $\mu$:
\bea
f_p(\bt,\mu) &=& \{\exp[\bt(E_p - \mu q_{5p})] +1\}^{-1},
\nonumber\\
E_p &=& |p|, \;\;\;\;
q_{5p} = p/|p|
\nonumber
\eea
(the axial charge of a fermion depends on the sign of $p$). 
Fig.\ \ref{fig2} shows $\ln(N_{{\rm av}p}^{-1} - 1)$ versus $ap$ 
at various times.
We see linear behavior, 
$\ln(N_{{\rm av}p}^{-1} - 1) \approx \bt(t)[|p| \pm\mu(t)]$, 
suggesting local (in time) equilibrium.
\begin{figure}[t]
\epsfxsize 7.9cm
\centerline{\epsfbox{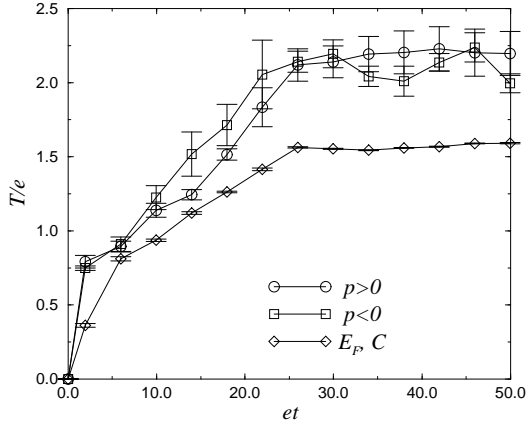}}

\vspace{-1cm}

\caption{Effective temperatures $T(t) = 1/\bt(t)$ versus time
obtained from fits as in figure 2.
}
\label{fig3}
\end{figure}
The distribution functions of modes with $ap \geq 0.5$ are consistent with zero,
so these modes are practically not excited. Furthermore, since the relevant $p$
are small in lattice units, discretization effects are reasonably small.
To achieve this the initial energy stored in the Bose fields has to be 
in the relatively low momentum modes only -- the fields are
far from classical equilibrium.

\begin{figure}[t]
\epsfxsize 7.9cm
\centerline{\epsfbox{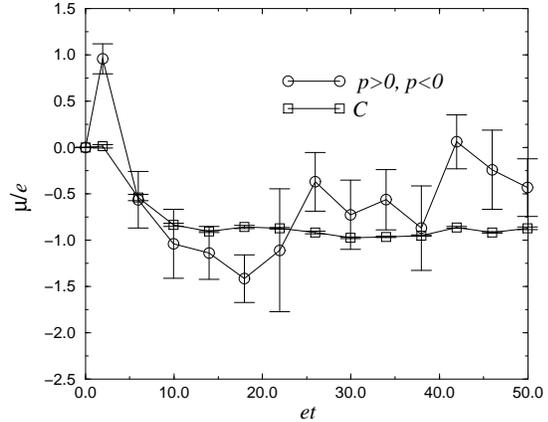}}

\vspace{-1cm}

\caption{As in figure 3 for the effective chemical potential.
}
\label{fig4}
\end{figure}

Figs.\ \ref{fig3} and \ref{fig4} show the effective temperature 
and chemical potential as a function of time. 
Note that $T_{p>0}(t) \approx T_{p<0}(t)$.
If the fermions were free, then their energy and axial charge
densities would follow from
the Fermi-Dirac distribution according to
\bea
E_F/L &=& n_f(\pi T^2/6 + \mu^2/2\pi),\nonumber\\
\langle Q_5\rangle/L &=& n_f \mu/\pi.\nonumber
\eea
Conversely, $E_F$ and 
$\langle Q_5\rangle$ 
imply an effective 
temperature and chemical potential; these are also plotted in Figs.\
\ref{fig3},\ref{fig4} (data labeled $E_F,C$ resp.\ $C$). 
The $E_F,C$-temperature appears systematically lower that
that from $N_{\rm avp}$. This may be due to the fact that 
the fermions are not free.
 
4. In conclusion, we see evidence for fast 
equilibration of fermions coupled to classical Bose fields. 
It is important
that the classical fields can be spatially inhomogeous, since this
allows the fermions to scatter nontrivially, e.g.\ by their
(screened) Coulomb interaction. The Bose fields have created fermions
and lost some energy, and they are not in (classical)
equilibrium in the time span shown here.
More details on this work can be found in \cite{AaSm99}.



\begin{thebibliography}{9}
\bibitem{Moore}     G.D.~Moore, these proceedings.
\bibitem{SB99}     G.\ Aarts, J.\ Smit,
        workshop on {\em Nonequilibrium Quantum Fields},
       Santa Barbara 1999, http://www.itp.ucsb.edu/online/noneq99.
\bibitem{AaSm98}     G.~Aarts and J.~Smit, Nucl.~Phys.~B555 (1999) 355.
\bibitem{AaSm99}     G.~Aarts and J.~Smit, hep-ph/9906538.
\end{thebibliography}
\end{document}